\documentclass[a4paper,11pt]{article}
\usepackage{pos}

\graphicspath :

\title{GRAND: status and perspectives}
 \ShortTitle{The GRAND Project}

\author*[a,b]{Kumiko Kotera}
\onbehalf{for the GRAND Collaboration}

\affiliation[a]{Sorbonne Universit\'{e} et CNRS, UMR 7095, Institut d'Astrophysique de Paris, 98 bis bd Arago, 75014 Paris, France}

\affiliation[b]{Vrije Universiteit Brussel (VUB), Dienst ELEM, Pleinlaan 2, B-1050, Brussels, Belgium
}

\affiliation[c]{Department of Physics; Department of Astronomy \& Astrophysics; Center for Multimessenger Astrophysics, Institute for Gravitation and the Cosmos, The Pennsylvania State University, University Park, PA 16802, USA}

\emailAdd{kotera@iap.fr}

\abstract{GRAND (the Giant Radio Array for Neutrino Detection) is a proposed next-generation observatory of ultra-high-energy neutrinos, cosmic rays, and gamma rays of cosmic origin, with energies exceeding about 100 PeV. GRAND is envisioned as a collection of large-scale ground arrays of self-triggered radio antennas that target the radio emission from extensive air showers initiated by UHE particles. Three prototype arrays are in operation: GRAND@Nançay in France, GRAND@Auger in Argentina, and GRANDProto300 in China. They test the detection principle and technology of GRAND, in preparation for its next phase, consisting of two arrays of 10'000 antennas each, in the Northern and Southern hemispheres, to be deployed from 2030 on. We present the concept of GRAND, its science goals, the status of the prototypes, their first measurements, and the technical and scientific perspectives that these measurements open for the field.}

\FullConference{10th International Workshop on Acoustic and Radio EeV Neutrino Detection Activities (ARENA)\\
		June 11 -- 14, 2024\\
		KICP, Chicago, USA}

\begin{document}
\maketitle

GRAND is a proposed large-scale observatory designed to unveil the most powerful sources in the Universe, by collecting their astroparticle multi-messengers  at ultra-high-energies (UHE). GRAND will detect the radio signals made in the Earth's atmosphere by UHE cosmic rays, gamma rays, and neutrinos. The combination of its high sensitivity, full-sky daily field of view, and sub-degree angular resolution will make possible the launch of UHE neutrino astronomy. GRAND is also foreseen to bring new insight to other science fields such as solar physics and transient radio astronomy. We present the detection concept and expected performances, the prototyping status, as well as the next steps to achieve the ultimate array.

\section{GRAND Concept}

\subsection{Detection concept and expected performances}

GRAND \cite{GRAND20, 2023arXiv230800120G} is designed to detect inclined extensive air-showers produced by UHE cosmic rays, gamma rays, and by Earth-skimming tau neutrinos that interact in the ground, producing tau particles, which in turn can emerge in the atmosphere generating cascades. The electromagnetic signal emitted mainly by the deflection of charged particles in these showers by the geomagnetic field can be observed by radio antenna arrays. 

In its final configuration, GRAND plans to deploy $200,000$ antennas over $200,000\,{\rm km}^2$, split into $\sim 20$ sub-arrays of $10,000$ antennas located in different locations across the Earth, in radio-quiet environments with easy access, and favorable topographies. Simulations show that ground topographies inclined by few degrees improve the detection efficiencies of a factor of three compared to flat ground \cite{Decoene_2021}.

\begin{figure}[t]
\centering
\includegraphics[height=0.4\linewidth]{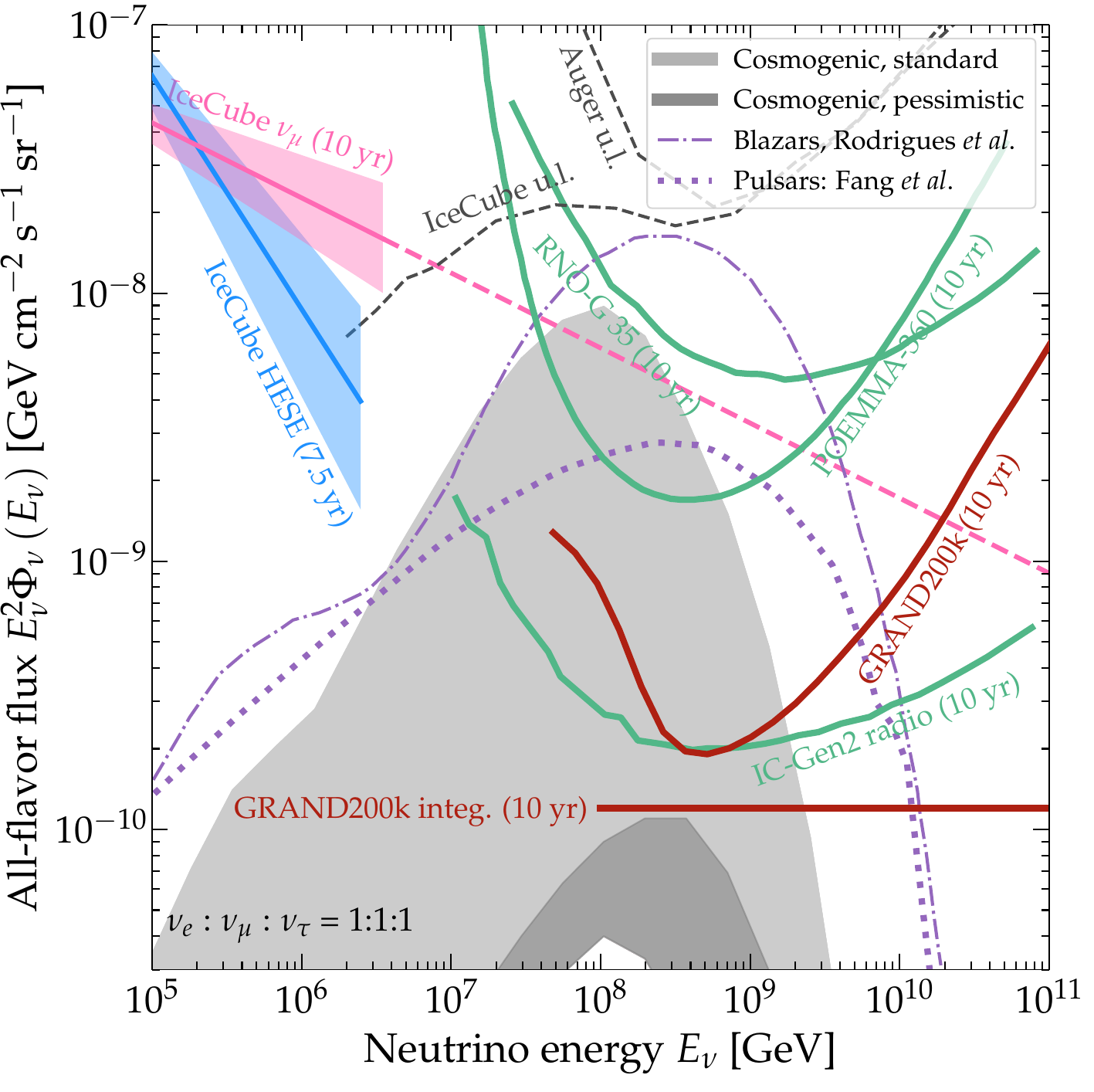}
\includegraphics[height=0.4\linewidth]{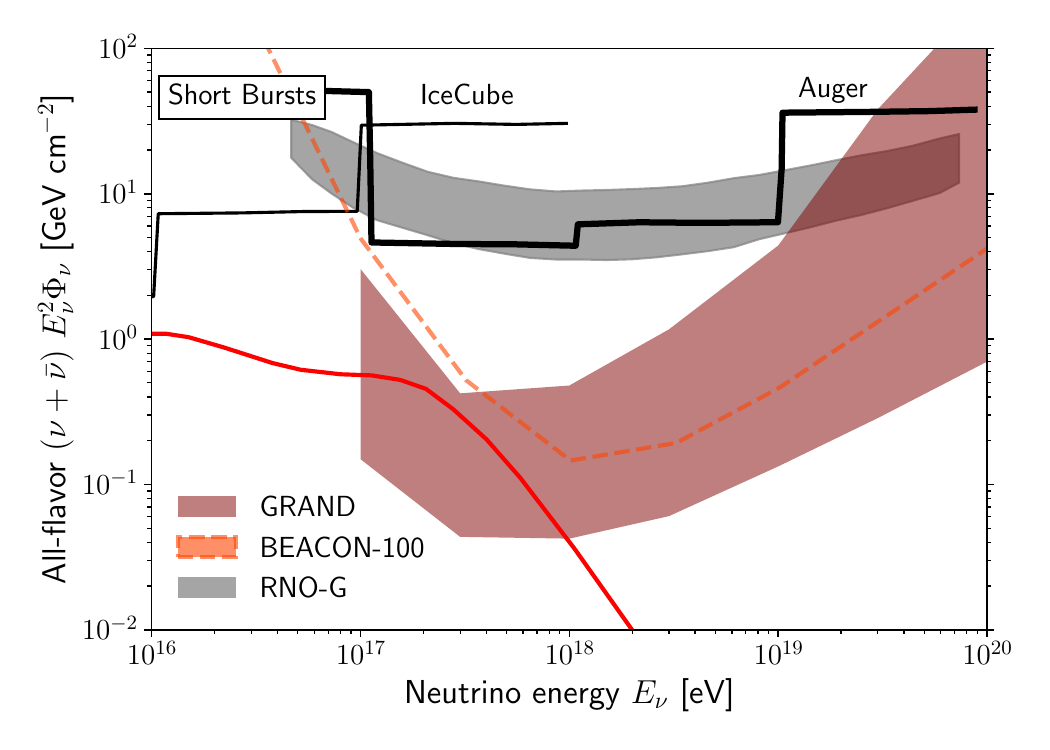}
\caption{{\it Left:} Projected 10-year sensitivities of GRAND and other projects (differential and integrated over an $E^{-2}$ spectrum, with $E$ the neutrino energy). Overlayed are predicted diffuse neutrino fluxes from astrophysical sources \cite{2014PhRvD..90j3005F,Rodrigues:2018bjg} (purple), of cosmogenic origin: standard (light gray band) and pessimistic (dark gray band) parameters~\cite{AlvesBatista:2018zui} and a theoretical extension to UHE energies of the measured IceCube flux, 
as well as upper limits on UHE neutrinos from IceCube 
and Auger. 
{\it Right:} GRAND instantaneous sensitivity over all declinations, assuming that the source is in the field of view in terms of right ascension, for 200k antennas deployed in a single location, are compared with the fluence of 100 stacked short GRBs or binary neutron star mergers within 100 Gpc \cite{Kimura_2017} (red solid line). The same quantity is indicated for RNO-G and BEACON 100 stations (adapted from \cite{2022NatRP...4..697G}).
}
\vspace{-0.3cm}
\label{fig:nu_performances}
\end{figure}

By construction, the instantaneous field of view of each GRAND sub-array for neutrinos is reduced to $\sim 6^\circ$ below the horizon, for a $360^\circ$ azimuthal view, corresponding to $\sim 6\%$ of the sky per site. The total instantaneous field of view hence scales roughly with the number of sub-arrays. With sites located in both hemispheres, the day-averaged sky coverage reaches 100\%. The instrument will reach $10^{-1}\,{\rm GeV\,cm}^{-2}$  instantaneous point source sensitivity (at zenith angle $90^\circ$) for neutrino energies $\sim 10^{18}$\,eV, and a 10-year integrated diffuse sensitivity limit $\sim 10^{-10}\,{\rm GeV}\,{\rm cm}^{-2}\,{\rm s}^{-1}\,{\rm sr}^{-1}$ above $5\times 10^{17}$\,eV (Fig.~\ref{fig:nu_performances}). Thanks to the kilometer-size radio footprints on the ground, an exquisite sub-degree angular resolution is expected to be reached over the accessible energy range~\cite{2023APh...14502779D}.

\subsection{Radio signals at our detectors}

GRAND will observe the coherent radio emission produced by UHE particles in the $50-200$\,MHz frequency range. The amplitude of the radiated electric field scales linearly with the particle energy. At each antenna, the signal corresponds to a pulse of duration $\lesssim 100\,$ns that triggers the acquisition system when a threshold of $3-5\,\sigma$ above the stationary Galactic background is reached. The radio emission can be considered point-like and distant, well represented by a spherical wave front around the shower axis. 

Because of relativistic effects, the radio emission is strongly beamed forward, with an opening angle of few degrees. Due to the propagation distance, the radio footprint on the ground can span several tens of kilometers. At frequencies $\gtrsim 100\,$MHz, a Cherenkov ring appears, due to the geometrical boost of the emission associated to the atmospheric refractive index. This feature is particularly useful for reconstruction procedure.

\subsection{GRAND technical challenges}

Due to its large scale, GRAND faces several technical challenges that are being addressed, experimented and tested with our prototypes.

\vspace{0.2cm}

\noindent {\bf Low-complexity, robust, and low-cost detection units} have to be designed and produced. The system needs to integrate an adequate shielding system to prevent electronics noise and be robust to extreme desert environments. 
The units have to be conceived for deployment in series. See \cite{proc_chiche_2024}.

\vspace{0.2cm}

\noindent {\bf Autonomous radio-detection}, i.e., identifying air-shower signals with radio antennas alone, will enable cost-efficient detection over large areas. It is a major challenge due to the ubiquitous radio background, necessitating an important rejection efficiency. The development of dedicated data acquisition (DAQ) electronics is required at antenna level, to enable high sampling rate ($\sim$\,kHz) and self-triggering. Furthermore, signal identification methods are being developed, based on previous studies on time traces, 
amplitude, 
and polarization patterns at ground, 
and experimental efforts in TREND \cite{CHARRIER201915}. 
These methods are being refined, and novel ideas and sophisticated data treatments (adaptative filtering, machine learning, etc.) are being developed \cite{Chiche_2022, 2018arXiv180901934F}. See \cite{proc_correa_2024, proc_koehler_2024}).

\vspace{0.2cm}

\noindent {\bf Data volume and transfer: low rate and low power consumption}. In its current design, GRAND will have to manage huge data volumes ($\sim 10\,$kBy/trigger). For the GRANDProto300 protype, the nominal trigger rates are: 1\,kHz for a basic threshold level (L1), and 10\,Hz for the CPU-based computation using template fitting or machine learning techniques (L2), at detection unit. The goal of the next phase of GRAND is to reduce these rates down to 100\,Hz for L1 and 1\,Hz for L2. Offline treatments demonstrate that the relevant information to identify an event from the background can be reduced to few quantities, and simple identification recipes can be implemented online. See \cite{proc_correa_2024, proc_koehler_2024}).

\vspace{0.2cm}

\noindent {\bf Reconstruction of very inclined shower parameters} is to be reinvented, as the geometries, physics at play for emission and propagation of these showers and their associated radio emissions differ drastically from the vertical showers studied previously \cite{Chiche_2024, Guelfand_2024}. See also \cite{proc_macias_2024}.

\section{Prototyping}

The GRAND collaboration has started the deployment of 3 complementary prototypes in 3 locations that will test  various aspects of the solutions developed to address the challenges above. In the rest of this proceeding, we will focus mainly on GRAND@Auger and GRANDProto300, which are large enough to collect statistics for technical validation and/or physics results.  

\paragraph{GRAND@Nançay} in central France, is a local test-bench of 4 antennas, to run hardware and trigger tests, deployed in fall 2022. 

\paragraph{GRAND@Auger} consists in 10 antennas deployed between March to November 2023 in Malargüe, Argentina, at the location of the Pierre Auger Observatory, using the AERA infrastructure. It will evaluate the quality of the GRAND reconstruction procedures, in terms of arrival direction, energy and nature of the primary particle. With an expected rate of around 1 cosmic ray shower/day in coincidence with Auger surface detector (SD) data, calibration and validation of the reconstruction event by event will be possible.

\paragraph{GRANDProto300 (GP300):} 13 out of the final 300 detection units (with positions spread over 200\,km$^2$ approved by the local authorities) were deployed at Xiao Dushan, Dunhuang, China, in February 2023, followed with a year of commissioning. The deployment of 76 antennas to complement the array will start fall 2024. This mid-scale prototype will validate the major challenges of GRAND. See also \cite{proc_chiche_2024}.

\subsection{Prototype set-up}

\begin{figure}[t]
\centering
\includegraphics[height=0.38\linewidth]{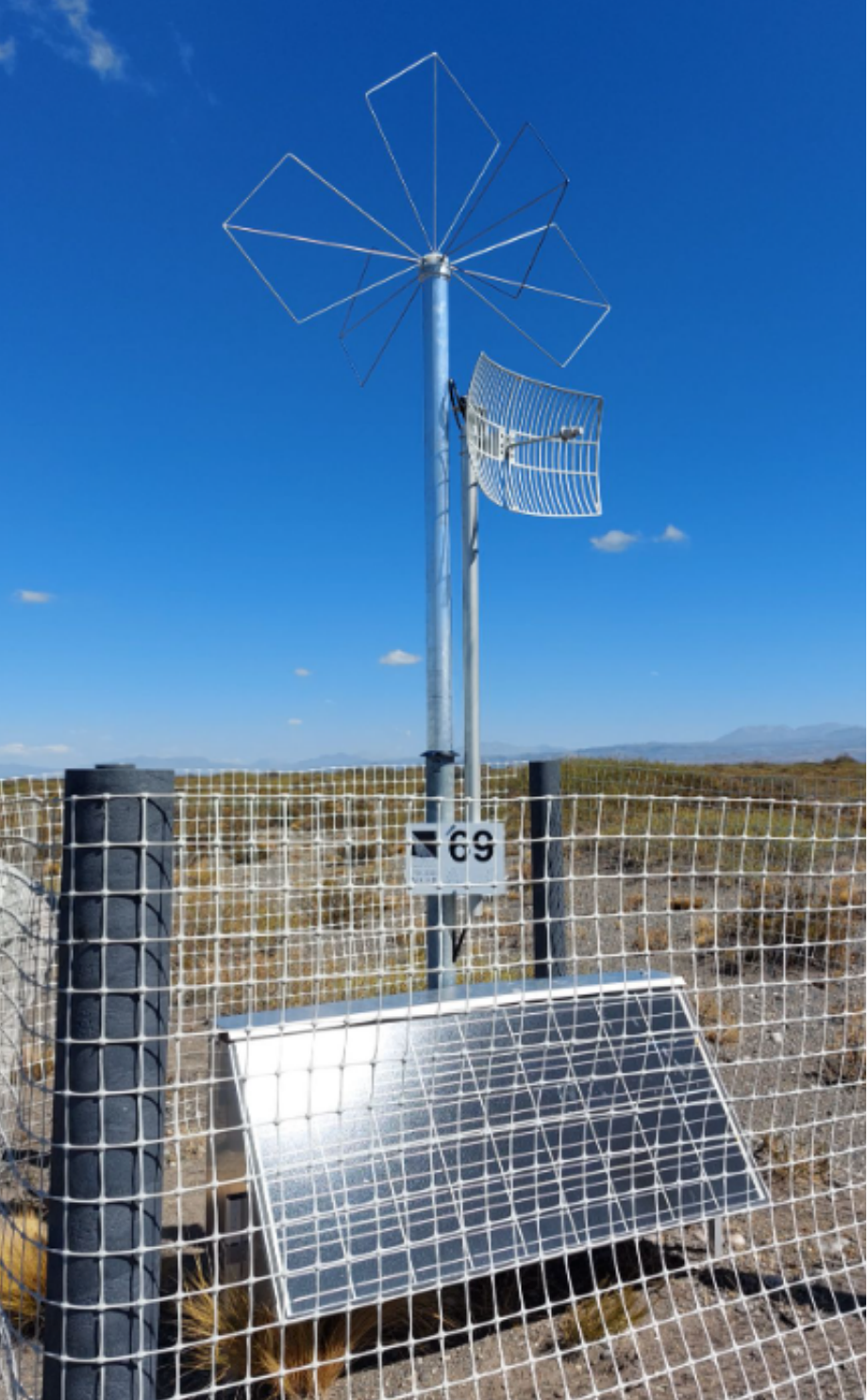}
\includegraphics[height=0.38\linewidth]{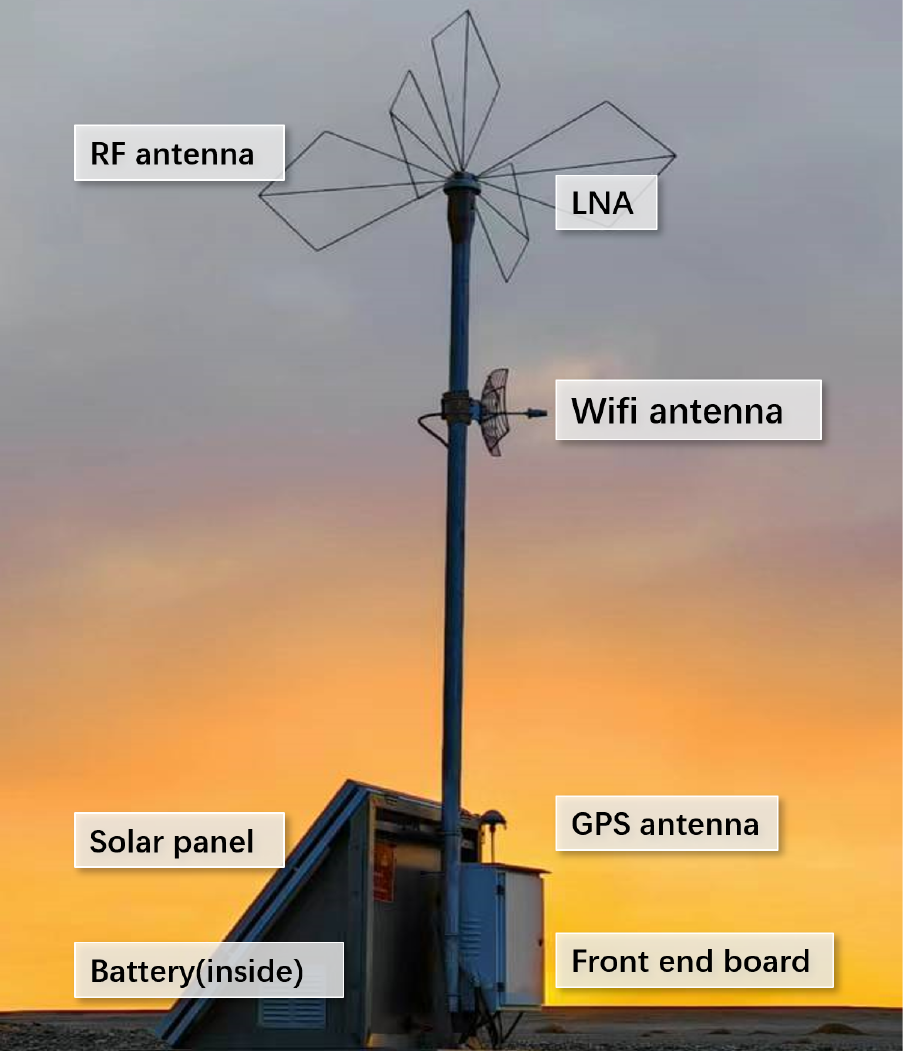}
\caption{GRAND@Auger (left) and GP300 (right) prototype detection units.}
\vspace{-0.3cm}
\label{fig:du_pic}
\end{figure}

GRAND@Auger and GP300 hardware set-ups present similar designs, with identical major components, central to the GRAND detector (Fig.~\ref{fig:du_pic}). However, several features are tuned to the requirements of the environment. The two prototypes also offer the possibility to test various options for specific components of the detector. 
Both prototype units are composed of a butterfly antenna, called HorizonAntenna, of 1.37 m extension. The gain pattern is maximal towards the horizon for very inclined shower detection, although the response of the antennas differs between GRAND@Auger and GP300 due to the mechanical structure supporting them. The Low Noise Amplifier (LNA) on which the antenna arms are plugged differs between the two set-ups to adapt to the more noisy environment of the AERA location. The LNA is mounted on top of a 3.5\,m pole, at the bottom which both structures harbor a triangular box closed by a solar panel, that contains the front-end board and the charge controller in Faraday-cage boxes, and the battery. In GP300, this container is placed on the ground, and serves as a weight for the unit. At GRAND@Auger, the AERA base has been conserved, and the DAQ triangular box is attached to the pole, at higher level from the ground. Data is transferred to the central DAQ via bullet wifi. The Front-End Board (traces recorded in Analog to Digital Convertor -- ADC -- units, with 500 M Samples/s, over 14 bits, with FPGA and 4 CPUs) is identical for the 2 set-ups. The trigger algorithm (unbiased trigger, ten second samples, 20 Hz mode) and the data format are also common. Two different firmware versions are implemented to test transient triggering. 

In terms of robustness to the environment, in Malargüe, the detection units had to face humidity and noise level issues, and accommodate constraints due to the mechanical structure in place, and the low-power consumption necessities. In  XiaoDushan, temperature fluctuations led to component overheating.

\subsection{First measurements and sets of reconstructed events}

\begin{figure}[t]
\centering
\includegraphics[width=0.49\linewidth]{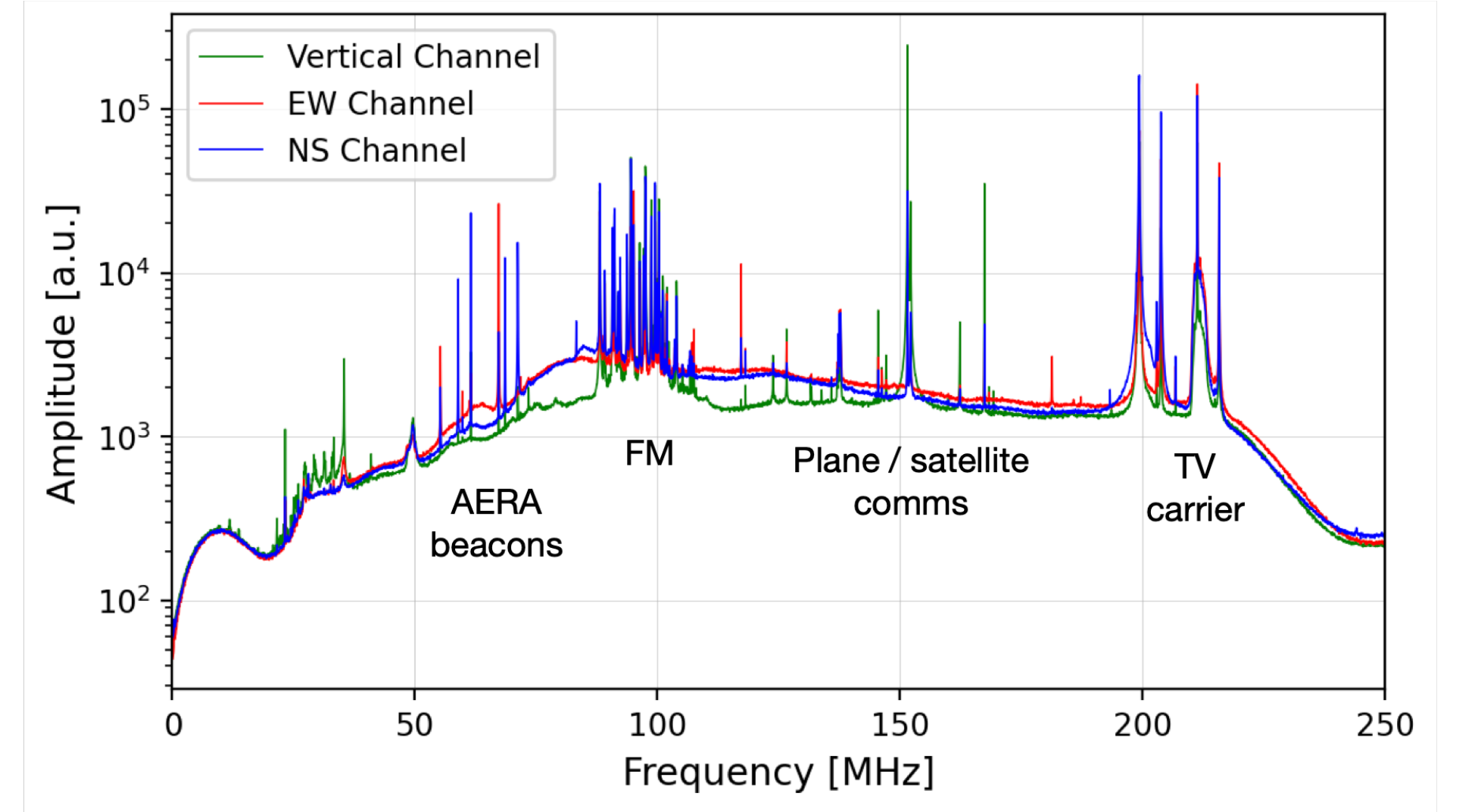}
\includegraphics[width=0.49\linewidth]{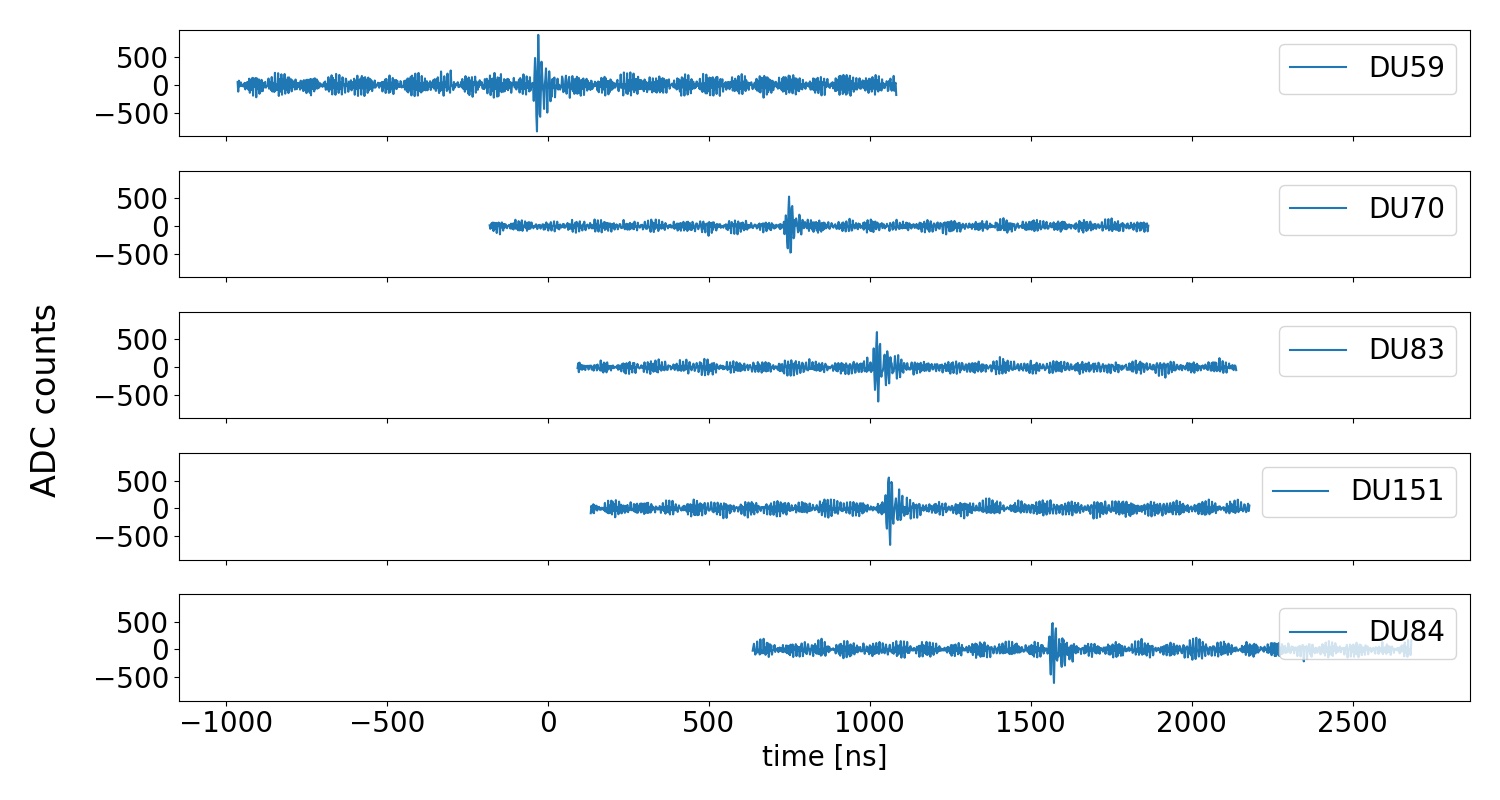}
\caption{{\it Left:} Power spectrum density measured with the GRAND@Auger prototype, along the 3 antenna direction channels, in arbitrary units. Labels indicate the identified peak sources. {\it Right:} Triggered ADC traces for a time-coincident event detected online with GRAND@Auger on 5 detection units (identities indicated by the labels). For similar plots for GP300, see \cite{proc_chiche_2024}.
}
\vspace{-0.3cm}
\label{fig:PSD_coinc}
\end{figure}

The 10 detection units of GRAND@Auger and the 13 units at GP300 have been taking data routinely since  November 2023. Preliminary measurements of the power spectrum density indicates a clean continuous background at GP300. At GRAND@Auger, most of the peaks observed in the spectrum are of identified human origin (Fig.~\ref{fig:PSD_coinc}, left). On both set-ups, coincident transients between different channels (antenna arms) and different detection units have been observed, either with online (GRAND@Auger) or with offline (GP300) searches (Fig.~\ref{fig:PSD_coinc}, right). On both prototypes, hardware tests on long-term stability, self-made noise control, LNA optimization, as well as firmware tests with trigger and measurements of transient rates are being performed. 

First source reconstructions have been performed on GRAND@Auger with online triggered coincident events (L1 trigger at detection unit level, and L3 trigger at central DAQ) and on GP300 with offline coincident searches (after L1 trigger at the detection unit). At GP300, the direction of the beacon (a 70\,MHz sine wave emitter), installed for calibration, was successfully reconstructed with an analytical Plane Wave Front (PWF) model \cite{Ferriere24} and a Spherical Wave Front (SWF) model \cite{CHARRIER201915}, with standard deviation of 10\,m on Northing and Westing, for a source located at 300\,m. More details on this procedure can be found in \cite{proc_chiche_2024}. The excellent position reconstruction is enabled by a timing precision of order 6\,ns on average over $\sim 15\,$min data collection. 

\begin{figure}[t]
\centering
\includegraphics[width=0.48\linewidth]{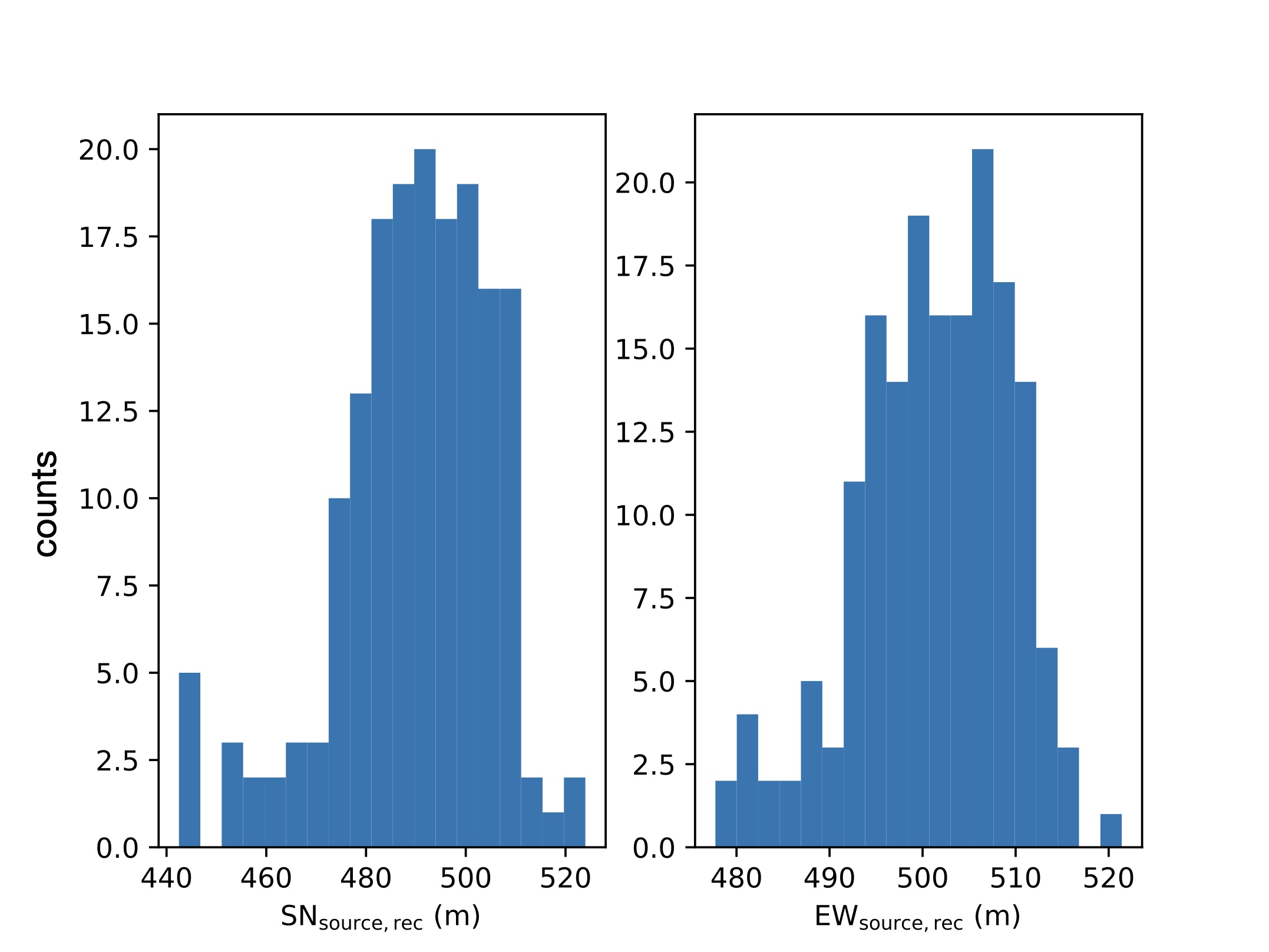}
\includegraphics[width=0.48\linewidth]{
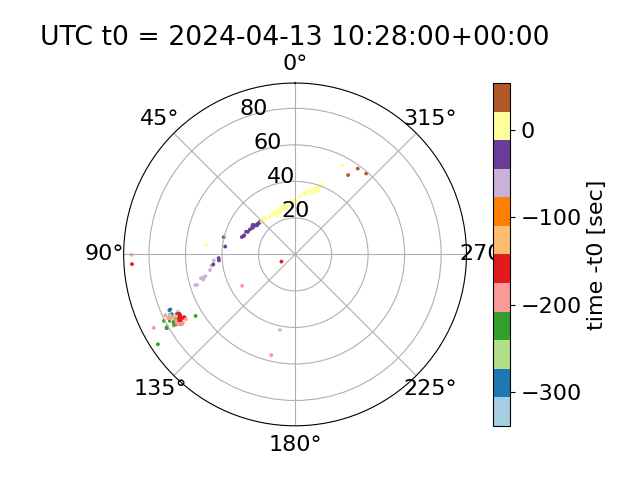}
\caption{First sets of reconstructed events with GP300 (offline coincidence search) and GRAND@Auger (online coincidence search at central DAQ, L3 trigger). {\it Left:} Reconstructed Northing and Westing positions of the beacon installed at GP300, with a SWF model (171 events reconstructed over 173 pulses emitted in the search time window of 10 minutes), with 10 m standard deviation. {\it Right:} Example of identified plane track over GRAND@Auger (TAM8133 from Santiago to Saõ Paulo) reconstructed with the analytical PWF method \cite{Ferriere24}. The color bar indicate the times relative to the registered trigger time stamp. See also \cite{proc_chiche_2024}). 
}
\vspace{-0.3cm}
\label{fig:recons_ev}
\end{figure}

At GRAND@Auger, 3 independent analyses  (analytic PWF \cite{Ferriere24}, Minuit PWF/SWF, TREND SWF \cite{CHARRIER201915}) consistently converge towards sources located in the direction of 2 villages nearby the site (Fig.~\ref{fig:recons_ev}, left). The events are based on online coincident searches (L3) between 3 to 7 detection units. Several plane tracks have also been identified, with time stamp correspondence to actual flights (Fig.~\ref{fig:recons_ev}, right). These events are being further analyzed to calculate transient detection rates and timing precision.

\section{Collaboration tools}

The GRAND Collaboration has been developing various tools to be ready to analyze the mid-scale prototype (80 antenna) phase of GP300, from 2025. 

\paragraph{Data flow and monitoring.} All GRAND raw data and simulations from ZHAiReS and CoREAS microscopic codes are converted into GRAND ROOT format, and stored at the CC-IN2P3 data center in Lyon. All runs and data files are logged in a database, and an online monitoring system enables the vizualisation of various features of the detection units and global set-ups at GP300 and GRAND@Auger (battery and temperature levels, signal RMS levels, traces and frequency domains, transient rate, coincidences).

\paragraph{Software pipeline: GRANDlib.} The collaboration has developed a software package, GRANDlib \cite{GRANDlibpaper}, designed to manage and analyze radio data from GRAND. It can be adapted for application in other experiments with large-scale astroparticle radio detection arrays. The primary purpose of GRANDlib is to generate simulations and to store and analyze data. The package handles terrestrial coordinate systems, Earth topography, and geomagnetic field. It is used to generate realistic radio signal templates from Monte-Carlo simulations for which it computes antenna response, radio-frequency (RF) chain parameters, and galactic noise. The package also has the capability to add other forms of noise, such as electronic noise, into the signal. GRAND data are stored in the ROOT file format, and GRANDlib includes a tool to manage and visualize them.

\paragraph{GRAND Reconstruction efforts.} New reconstruction methods have to be developed for the showers to be detected by the GRAND prototypes. The collaboration has built a realistic library of radio simulations to assess reconstruction performances (more than 200,000 simulations, raw and hardware like, available both in ADC and electric field trace formats, including antenna response and RF chain with jitter mimicking Galactic noise, trigger time and amplitude calibration smearings). 
This library also serves for triggering studies (at L2 and L3 levels, see \cite{proc_correa_2024, proc_koehler_2024}).

The various methods developed and tested are twofold: 1) reconstruction of the electric field from the data: with CNN, polarization methods, denoising using Machine Learning, and 2) shower parameter reconstruction: analytical PWF (comprising analytical error estimate) \cite{Ferriere24}, improved fitting (empirical and Physics informed) of Angular Distribution Function (ADF) based on \cite{2023APh...14502779D}, empirical fitting of lateral distribution function based on \cite{Schl_ter_2023}, GNN methods. A comprehensive review of these methods can be found in \cite{proc_macias_2024}.

\section{The road to neutrino astronomy}

The next step of GRAND will be the deployment of the 76 antennas of GP300, which will enable, already in 2025, to start validating the detection concept of the experiment. Meanwhile, physics cases will be studied, for the Galactic to extragalactic transition region of cosmic rays, and for Fast Radio Burst searches, to optimize the design of the remaining antennas to instrument the available 200\,km$^2$ region, with 220 additional units (to be deployed starting 2026). In parallel, the collaboration has started the R\&D for the next large-scale phase of GRAND, with exploratory simulations and testing of new designs in terms of hardware for each detection unit (for robustness and cost-efficiency), layout and concept. One possible option considered is a hybrid design that would extract the advantages of a phased radio array a la BEACON~\cite{Southall:2022yil} and the autonomous sparse array of GRAND, and place a large-scale array at high elevation \cite{proc_wissel_24}. Prototypes for the next phase of GRAND will be produced and tested until early 2030, when deployment should start, for a discovery instrument, with sensitivity to detect the first UHE neutrinos. 

By the 2030s, once this first sub-array has been demonstrated to operate successfully, its design will be duplicated with an industrial approach, with predefined specifications for reliability, costs etc. The design of each sub-array may be adapted, depending on location and topography, or to address specific science cases, the primary goal being UHE neutrino astronomy.

\bibliographystyle{elsarticle-num}
{\footnotesize
\bibliography{references}
}

\end{document}